\documentclass[aps,pra,twocolumn,showpacs,preprintnumbers,floatfix]{revtex4}
\usepackage{amsmath}
\usepackage{amssymb}
\usepackage{epsfig}

\begin{document}

\title{Quantum versus classical phase-locking transition in a driven-chirped
oscillator}
\author{I. Barth$^{^{1}}$, L. Friedland$^{^{1}}$, O. Gat$^{^{1}}$, and A.G.
Shagalov$^{^{2}}$}
\affiliation{$^{^{1}}$Racah Institute of Physics, Hebrew University of Jerusalem,
Jerusalem 91904, Israel \\
$^{^{2}}$Institute of Metal Physics, Ekaterinburg 620219, Russian Federation}

\begin{abstract}
Classical and quantum-mechanical phase locking transition in a nonlinear
oscillator driven by a chirped frequency perturbation is discussed.
Different limits are analyzed in terms of the dimensionless parameters $%
P_{1}=\varepsilon /\sqrt{2m\hbar \omega _{0}\alpha }$ and $P_{2}=(3\hbar
\beta )/(4m\sqrt{\alpha })$ ($\varepsilon ,$ $\alpha ,$ $\beta $ and $\omega
_{0}$ being the driving amplitude, the frequency chirp rate, the
nonlinearity parameter and the linear frequency of the oscillator). It is
shown that for $P_{2}\ll P_{1}+1$, the passage through the linear resonance
for $P_{1}$ above a threshold yields classical autoresonance (AR) in the
system, even when starting in a quantum ground state. In contrast, for $%
P_{2}\gg P_{1}+1$, the transition involves quantum-mechanical energy ladder
climbing (LC). The threshold for the phase-locking transition and its width
in $P_{1}$ in both AR and LC limits are calculated. The theoretical results
are tested by solving the Schrodinger equation in the energy basis and
illustrated via the Wigner function in phase space.
\end{abstract}

\pacs{42.50.Hz, 42.50.Lc, 33.80.Wz, 05.45.Xt}
\date{\today }
\maketitle

\section{Introduction}

Autoresonance (AR) is a generic nonlinear phase-locking phenomenon in
classical dynamics. It yields a robust approach to excitation and control of
nonlinear oscillatory systems by a continuous self-adjustment of systems'
parameters to maintain the resonance with chirped frequency perturbations.
Applications of AR exist in many fields of physics, examples being atomic
and molecular systems \cite{Maeda 2007,Chelkowski}, nonlinear optics \cite%
{Segev}, Josephson junctions \cite{Ofer Naaman}, hydrodynamics \cite%
{BenDavid}, plasmas \cite{Danielson}, nonlinear waves \cite{Lazar92}, and
quantum wells \cite{Manfredi 2007}. Most recently, AR served as an essential
element in the formation of trapped anti-hydrogen atoms at CERN \cite{ALPHA
Nature, ALPHA PRL} and in studying the effect of fluctuations in driven
Josephson junctions \cite{Kater}. While the classical AR is well understood,
the investigation of the quantum-mechanical limits of the problem has
started only recently \cite{Gilad, Manfredi 2007, Kater}. The present study
focuses on the interrelation between the classical and quantum descriptions
of the autoresonant transition in the simplest case of a driven Duffing
oscillator (modeling a driven diatomic molecule \cite{Zigler} or a Josephson
junction \cite{Ofer Naaman}, for example) governed by the Hamiltonian

\begin{equation}
H=\frac{p^{2}}{2m}+m\omega _{0}^{2}\left( \frac{1}{2}x^{2}-\frac{1}{4}\beta
x^{4}\right) +\varepsilon x\cos \varphi _{d},  \label{eq:Hamiltonian}
\end{equation}%
where $\varphi _{d}=\int \omega _{d}dt$, $\omega _{d}=\omega _{0}-\alpha t$
is the chirped driving frequency and $\alpha ,\beta >0$. We will assume that
initially our oscillator is in a thermal equilibrium with the environment at
temperature $T$, but the chirped system's response is sufficiently fast to
neglect the effect of the environment on the out-of-equilibrium dynamics
\cite{Kater}.

Classically, in autoresonance, after passage through the linear resonance at
$t=0$, the driven oscillator gradually self-adjusts its oscillation
frequency to that of the drive by continuously increasing its energy \cite%
{Fajans}, yielding a convenient control of the dynamics by variation of an
external parameter (the driving frequency). The transition to the classical
AR by passage through linear resonance has a threshold on the driving
amplitude, scaling as $\varepsilon ^{cr}\sim \beta ^{-1/2}\alpha ^{3/4}$
\cite{Fajans}. This threshold is sharp if the oscillator starts in its zero
equilibrium, but in the presence of thermal noise it develops a width,
scaling as $T{}^{1/2}$ \cite{PRL}. Both the AR threshold and its width have
their quantum-mechanical counterparts, which will be discussed in this work.

When the problem of autoresonant transition is dealt with
quantum-mechanically, two questions must be addressed. First, what are the
differences between the classical and quantum evolutions of the
chirped-driven nonlinear oscillator? In dealing with this question, Ref.
\cite{Gilad} suggested that the natural quantum-mechanical limit of the
classical AR is a series of successive Landau-Zener (LZ) \cite{LZ}
transitions or energy ladder climbing (LC), where only two adjacent energy
levels of the driven oscillator are coupled at any given time. In contrast,
the classical AR behavior takes place when many levels are coupled at all
times during the excitation \cite{Goggin}. We will adopt and further develop
this point of view here and describe different regimes in the problem in
terms of two dimensionless parameters $P_{1,2\text{ }}$suggested in \cite%
{Gilad}. These parameters are defined via the three physical time-scales in
the system, i.e. the inverse Rabi frequency $T_{R}=\sqrt{2m\hbar \omega _{0}}%
/\varepsilon $, the frequency sweep time scale $T_{S}=1/\sqrt{\alpha }$, and
the characteristic nonlinearity time scale $T_{NL}=(3\hbar \beta )/(4m\alpha
)$ (the time of passage through the nonlinear frequency shift between the
first two transitions on the energy ladder). Then, by definition

\begin{equation}
P_{1}=\frac{T_{S}}{T_{R}}=\frac{\varepsilon}{\sqrt{2m\hbar \omega _{0}\alpha
}}  \label{1}
\end{equation}
(this parameter measures the strength of the drive), and

\begin{equation}
P_{2}=\frac{T_{NL}}{T_{S}}=\frac{3\hbar \beta }{4m\sqrt{\alpha }}  \label{2}
\end{equation}%
(a measure of the nonlinearity in the problem). We will show in this work
that this parameter space describes all limiting cases of quantum-mechanical
evolution in our system, including quantum initial conditions, the
subsequent transition to either LC or AR, and the associated threshold
phenomenon. Note, that $P_{1,2\text{ }}$have a meaning only in the case of a
chirped system, because of the new time scale, $T_{S}$, associated with this
case.

The second question, which must be addressed in the quantum-mechanical
formulation of our problem is that of quantum fluctuations. As mentioned
above, in the presence of thermal noise, the classical AR transition
probability develops a width, scaling as $T{}^{1/2}$ with temperature \cite%
{PRL}. Nevertheless, at very low temperatures, the quantum fluctuations
should be taken into account. Recent experiments by Kater et al. \cite{Kater}
demonstrated quantum saturation of the width of the phase-locking transition
in superconducting Josephson junctions at sufficiently low temperatures,
confirming the prediction that $T$ in the classical width formula \cite{PRL}
should be replaced by an effective temperature, $T_{\mathrm{eff}}$, where $%
T_{\mathrm{eff}}=T$ for high temperatures and saturates at $T_{\mathrm{eff}%
}=\hbar \omega _{0}/2k_{B}$ at low temperatures. The experimental results
imply that the fluctuations only determine the initial conditions of such a
non-equilibrium oscillator and do not affect its time evolution. In this
work, we will address the effect of quantum \ fluctuations in the AR problem
theoretically and provide further justification of using the classical AR
threshold width formula with $T$ replaced by $T_{\mathrm{eff}}$.

The scope of the paper will be as follows. In Sec. II we will use the
quantum-mechanical energy basis in the rotating wave approximation and
compare the driven dynamics of our oscillator in the quantum and classical
regimes numerically. Section III will present the analytic description of
the transition to phase-locking in terms of the $P_{1,2}$ parameter space in
both classical AR and quantum LC regimes. In the same Section, the theory
will be compared with numerical simulations. Section IV will focus on the
effect of quantum fluctuations on the width of the phase-locking transition.
Finally, we will address the phase space dynamics in the problem in Sec. V
by solving the quantum Liouville equation for the Wigner function
numerically and compare the phase space evolution with that in the energy
basis. Our conclusions will be summarized in section VI.

\section{Chirped dynamics in the energy basis}

\begin{figure*}[tbp]
\includegraphics[width=18cm]{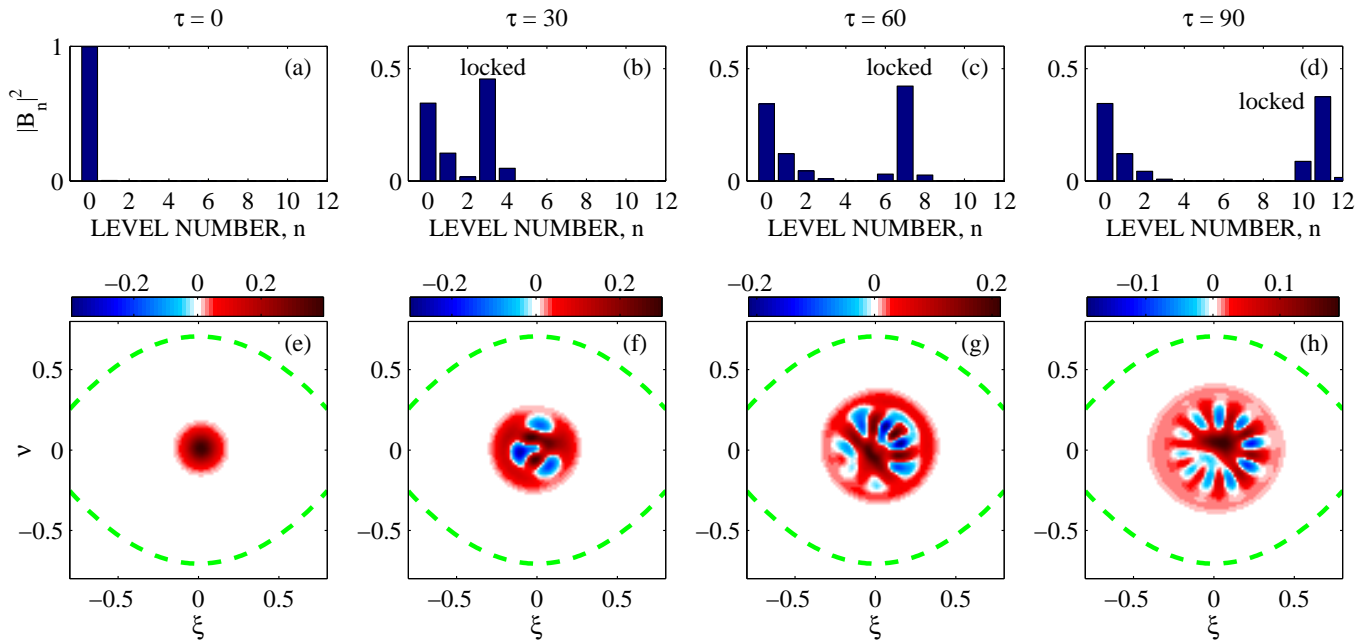}
\caption{(color online) The dynamics in the energy basis (a-d) and the
corresponding phase space dynamics of the Wigner function (e-h) in the
quantum ladder climbing regime, $P_{2}=8$. The subplots correspond to times $%
\protect\tau =0$ (a,e), $30$ (b,f), $60$ (c,g), and $90$ (d,h). Only a
single level is highly populated in the phase-locked group of levels. The
dashed lines in subplots (e-h) are the separatrices of the external
potential well. The dimensionless phase space coordinates are rescaled as $%
\protect\xi =\protect\sqrt{\overline{\protect\beta }}\bar{x}$, $\protect%
\upsilon = \protect\sqrt{\overline{\protect\beta }}\overline{u}$. }
\label{Flo:P2=8}
\end{figure*}

We write the wave function of the oscillator governed by Eq.(\ref%
{eq:Hamiltonian}), $|\psi\rangle=\sum_{n}c_{n}|\psi _{n}\rangle$, in the
energy basis $|\psi _{n}\rangle$ of the undriven ($\varepsilon =0$)
Hamiltonian (\ref{eq:Hamiltonian}). The associated Schrodinger equation
yields

\begin{equation}
i\hbar \frac{dc_{n}}{dt}=E_{n}c_{n}+\frac{\tilde{\varepsilon}}{\sqrt{2}}%
\left( \sqrt{n+1}c_{n+1}+\sqrt{n}c_{n-1}\right) \cos \varphi _{d},
\label{eq:schrodinger}
\end{equation}%
where we approximate the energy levels \cite{Landau QM}

\begin{equation}
E_{n}\approx \hbar \omega _{0}\left( n+1/2-\beta _{q}(n^{2}+n+1/2)\right) ,
\label{En}
\end{equation}%
$n=0,1,2,...$, $\beta _{q}=\frac{3\beta \hbar }{8m\omega _{0}}$, and $\tilde{%
\varepsilon}=\varepsilon \sqrt{\frac{\hbar }{m\omega _{0}}}$. We assume a
weak coupling, $\tilde{\varepsilon}\ll E_{0}$, and, consequently, neglect
the nonlinear correction of order $\tilde{\varepsilon}\beta _{q}/\hbar
\omega _{0}$ in the coupling term. Next, we define $C_{n}=e^{i\omega
_{n}t}c_{n}$, where $\omega _{n}=E_{n}/\hbar $, substitute this definition
into Eq. (\ref{eq:schrodinger}), and neglect the nonresonant terms (rotating
wave approximation) to get
\begin{eqnarray}
i\hbar \frac{dC_{n}}{dt} &\approx &\frac{\tilde{\varepsilon}}{2\sqrt{2}}(%
\sqrt{n+1}C_{n+1}e^{-i(\omega _{n,n+1}t-\varphi _{d})}  \notag \\
&&+\sqrt{n}C_{n-1}e^{i(\omega _{n-1,n}t-\varphi _{d})}),  \label{Cn}
\end{eqnarray}%
where $\omega _{n,n+1}=\omega _{n+1}-\omega _{n}=\omega _{0}-2\omega
_{0}(n+1)\beta _{q}$. Finally, we introduce $B_{n}=C_{n}e^{-i\int \gamma
_{n}dt}$, where $\gamma _{n}=n\alpha t-n(n+1)\omega _{0}\beta _{q}$ and the
dimensionless slow time $\tau =\sqrt{\alpha }t$, associated with the change $%
\tau ^{2}/2$ of the driving phase due to the driving frequency chirp. Then
Eq. (\ref{Cn}) can be written as

\begin{equation}
i\frac{dB_{n}}{d\tau }=\Gamma _{n}B_{n}+\frac{P_{1}}{2}\left( \sqrt{n+1}%
B_{n+1}+\sqrt{n}B_{n-1}\right) ,  \label{eq:slow eq.}
\end{equation}%
where $\Gamma _{n}=n[\tau -(n+1)P_{2}/2]$, and $P_{1}=\varepsilon /\sqrt{%
2\alpha \hbar \omega _{0}m},$ $P_{2}=2\omega _{0}\beta _{q}/\sqrt{\alpha }$,
as defined in the Introduction. Note, that $P_{1}$ characterizes the
strength of the coupling between the adjacent levels, while $P_{2}$ is
associated with the nonlinearity in the problem and determines the degree of
classicality in the system (see Sec. V). Note also that the rotating frame
here is \emph{chirped} instead of the usual, fixed frequency frame and,
thus, there remains an explicit time dependence in Eq. (\ref{eq:slow eq.}).
Our goal is to analyze these slow evolution equations, but first, we discuss
different limits in the driven system in $P_{1,2}$ parameter space.

The comparison between the classical AR and the quantum LC regimes was first
discussed by Marcus et al. \cite{Gilad}, who suggested the nonlinear
resonance classicality criterion, $P_{2}\ll P_{1}$, by requiring that the
classical resonance width would include more than two quantum levels. Since
the chirp rate cancels from this creterion, the latter characterizes the
nonlinear resonance phenomenon in the system driven by constant frequency
drive as well. The chirping introduces a new effect, i.e. a possibility of a
continuous self-adjustment of the energy of the oscillator to stay in
resonance with the drive. This yields a new condition, separating the
classical AR and quantum LC transitions, where the dynamics of the chirped
system is very different. In the LC transition, only two levels are coupled
at a time and the system's wave function climbs the energy ladder by
successive LZ transitions \cite{LZ}. For example, Eq. (\ref{eq:slow eq.})
yields the following two-level transformation matrix for the $n-1\rightarrow
n$ transition

\begin{equation}
\left(
\begin{array}{cc}
(n-1)\tau -\frac{n(n-1)}{2}P_{2} & \frac{\sqrt{n}}{2}P_{1} \\
\frac{\sqrt{n}}{2}P_{1} & n\tau -\frac{n(n+1)}{2}P_{2}%
\end{array}%
\right) .  \label{TrMatrix}
\end{equation}%
We can calculate the time of the $n^{th}$ transition, $\tau _{n}$ by
equating the diagonal elements in this matrix, i.e. $\tau _{n}=nP_{2}$, so
the time interval between two successive transitions is $\Delta \tau =P_{2}$%
. On the other hand, the typical duration $\Delta \tau _{LZ}$ of each LZ
transition has two distinct limits \cite{Finite LZ}. In the non adiabatic
(sudden) limit ($P_{1}\ll 1$), $\Delta \tau _{LZ}$ is of the order of unity,
while in the opposite (adiabatic) limit, $\Delta \tau _{LZ}\sim P_{1}$.
Therefore, by comparing $\Delta \tau $ and $\Delta \tau _{LZ}$, we expect to
see well separated successive LZ steps, i.e. the LC, provided $P_{2}\gg
P_{1}+1$ which describes both the sudden and the adiabatic limits. In
contrast, the classical AR transition requires $P_{2}\ll P_{1}+1$, which
coincides with the nonlinear resonance classicality criterion mentioned
above, when $P_{1}\gg 1$. In section V, a different argument will be
suggested to explain why classical mechanics yields the correct description
of the transition to autoresonance when a stronger inequality, $P_{2}\ll 1$,
is satisfied, even when the system starts in the quantum mechanical ground
state. Next, we discuss the numerical solutions of the problem and compare
different regimes of chirped-driven dynamics.

\begin{figure*}[tbp]
\includegraphics[width=18cm]{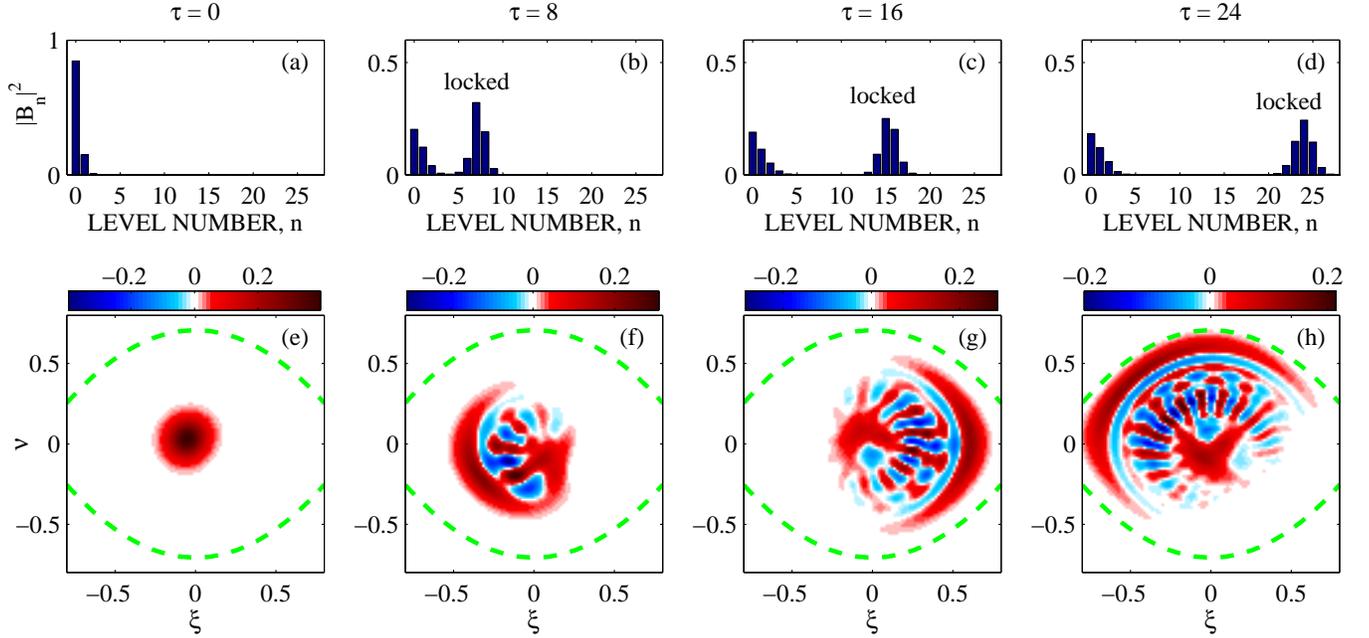}
\caption{(color online) The dynamics in the energy basis (a-d) and the
corresponding phase space dynamics of the Wigner function (e-h) in the
intermediate regime, $P_{2}=1$. The subplots correspond to times $\protect%
\tau=0$ (a,e), $8$ (b,f), $16$ (c,g), and $24$ (d,h). Few levels are
simultaneously excited in the phase-locked group in the intermediate regime.
The dashed lines in (e-h) are the separatrices of the external potential
well. The dimensionless phase space coordinates are rescaled as $\protect\xi %
=\protect\sqrt{\overline{\protect\beta }}\bar{x}$, $\protect\upsilon =
\protect\sqrt{\overline{\protect\beta }}\overline{u}$.}
\label{Flo:P2_1}
\end{figure*}

\begin{figure*}[tbp]
\includegraphics[width=18cm]{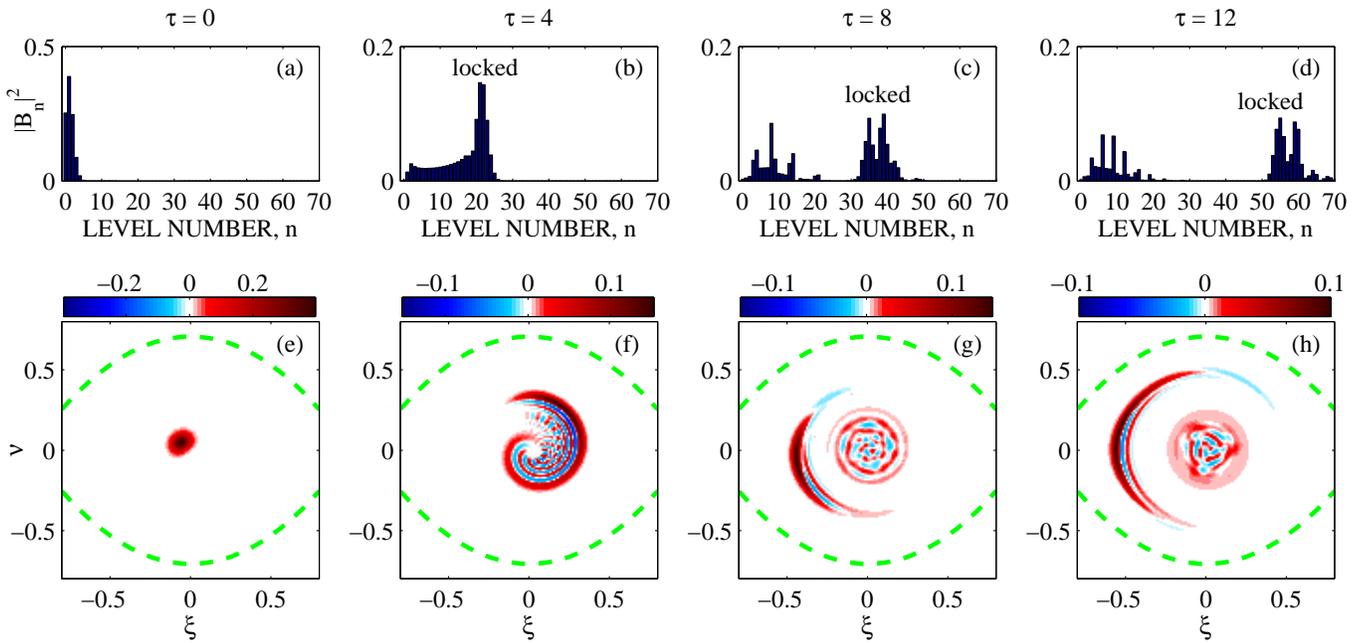}
\caption{(color online) The dynamics in the energy basis (a-d) and the
corresponding phase space dynamics of the Wigner function (e-h) in the
classical AR regime, $P_{2}=0.2$. The subplots correspond to times $\protect%
\tau=0$ (a,e), $4$ (b,f), $8$ (c,g), and $12$ (d,h). Many levels are excited
in the phase-locked group. The dashed lines in (e-h) are the separatrices of
the external potential well. The dimensionless phase space coordinates are
rescaled as $\protect\xi =\protect\sqrt{\overline{\protect\beta }}\bar{x}$, $%
\protect\upsilon = \protect\sqrt{\overline{\protect\beta }}\overline{u}$.}
\label{Flo:P2=0.2}
\end{figure*}

We have solved Eqs. (\ref{eq:slow eq.}) numerically, subject to ground state
initial conditions $B_{n}(\tau _{0})=\delta _{n,0}$ at $\tau _{0}=-8$ (the
linear resonance corresponds to $\tau =0$). Each of the Figs. \ref{Flo:P2=8}%
--\ref{Flo:P2=0.2} corresponds to a different value of the nonlinearity
parameter $P_{2}$ and show the distribution of the population of the levels
in the system at four different times (subplots a-d). The subplots e-h in
the Figures show the associated Wigner distributions (see Sec. V) at the
same times. Figure \ref{Flo:P2=8}\ shows the case of the LC dynamics for $%
P_{2}=8$ and $P_{1}=0.8$ at $\tau =0,\,30,\,60$, and $90\ $(subplots a-d),
and illustrates a clear time separation beyond the linear resonance between
the successive LZ transitions. For example, we observe two groups of
resonant and nonresonant levels at $\tau =90$, separated by a valley
centered at about $n=6$. We find that the location of the resonant levels is
determined by the slow time, i.e. $n\approx \tau /P_{2}$, as shown above.
Thus, the resonant (phase-locked) state in the system is efficiently
controlled via the driving frequency and a given final state can be reached
(and maintained) by terminating the frequency chirp at the desired energy
level. We also see that there exists a single highly occupied level in the
resonant group of levels at any given time, indicating successive LZ
transitions, as expected in the LC regime.

Our second numerical example is presented in Fig. \ref{Flo:P2_1} and
illustrates the intermediate regime (as discussed above) with $P_{1}=P_{2}=1$
and $\tau =0,8,16$, and $24$ (the subplots a-d). As in Fig. \ref{Flo:P2=8},
a clear separation between the resonant $(n<5)$ and nonresonant $(n>20)$
groups of levels is seen in the Figure. We see that, typically, several
levels are excited in the resonant group, but their number is small, so the
driven dynamics can not be considered as classical. The last example (see
Fig. \ref{Flo:P2=0.2}) corresponds to the classical regime, $P_{2}=0.2$, $%
P_{1}=1.9$ and $\tau =0,4,8,12$. One observes a separation between resonant
and nonresonant groups at $\tau =12$. Note that in all our numerical
examples about 50\% only of the initial state is transferred to the
continuing phase-locked state, leading to the question of resonant capture
probability, which is discussed next.

\section{Resonant capture probability}

\subsection{Threshold for phase-locking transitions}

For a given set $(P_{1},P_{2})$ we define the resonant capture probability,

\begin{equation}
P=\sum_{n=n_{c}}^{\infty }\left\vert B_{n}\right\vert ^{2}
\label{eq:Numeric_Prob}
\end{equation}%
where $n_{c}$ is the number of the level separating the resonant and
nonresonant groups of levels at sufficiently large times. For a given value
of $P_{2}$, the probability $P$ depends on the driving parameter, $P_{1}.$
For example, in the case in Fig. \ref{Flo:P2=8}, we use $n_{c}=6$ and the
resonant capture probability is $P=0.48$. Similarly, in the two examples in
Figs. 2 and 3, we choose $n_{c}=10,\ 40$ to get $P=0.62,\ 0.66$,
respectively.

\begin{figure}[tbp]
\centering\includegraphics[width=8cm]{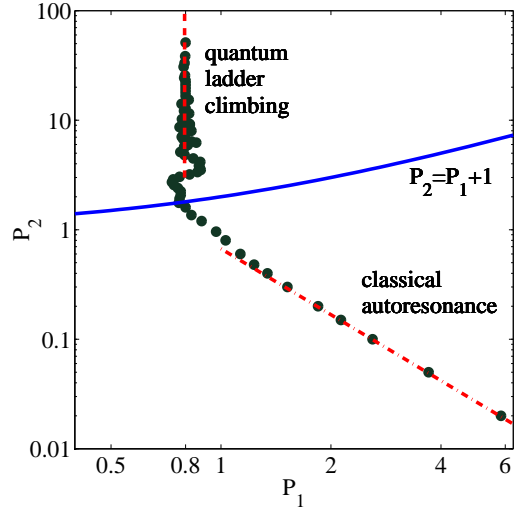}
\caption{(color online) Different regimes of phase-locking transition in the
chirped oscillator. The full circles show the location of the numerical $1/2$
resonant capture probability (the threshold for phase-locking transition)
obtained by solving the Schrodinger Eq. (\protect\ref{eq:slow eq.}) subject
to initial condition in the ground state. The dashed and dashed-dotted lines
represent the theoretical thresholds in the LC and AR regimes, respectively.
The line $P_2=P_1+1$ separates the classical AR and the quantum LC regimes.}
\label{Flo:center}
\end{figure}

We calculate the resonant capture probability by solving Eqs. (\ref{eq:slow
eq.}) numerically subject to initial conditions, $B_{n}(\tau _{0})=\delta
_{n,0}$ (the ground state), for different values of $P_{1,2}$ and $\tau
_{0}=-10$. For a fixed $P_{2}$, the capture probability $P$ is a
monotonically increasing, smoothed step function of $P_{1}$. We define the
threshold for \textit{efficient} phase-locking transition, $P_{1}^{cr}$, as
the value of $P_{1}$ for 1/2 capture probability, i.e. $P(P_{1}^{cr})=0.5$.
The full circles in Fig. \ref{Flo:center} show $P_{1}^{cr}$ for different
values of $P_{2}$. The dashed and dashed-dotted lines are the assymptotic
theoretical predictions for the quantum LC and classical AR (see below),
which agree with the results of our simulations in both limits. The line $%
P_{2}=P_{1}+1$ is the separator between the classical and the quantum
regimes of the chirped nonlinear resonance, as discussed in Sec. II. This
line crosses the threshold line $P_{1}^{cr}$ at $(P_{1},P_{2})\approx
(0.8,1.8)$. One can see in the Figure that indeed, this point separates very
different dependences of $P_{1}^{cr}$ on $P_{2}~$associated with the quantum
and classical dynamics of the chirped system. One can also see the
oscillating pattern of the threshold $P_{1}^{cr}$ at $1<P_{2}<5$, where the
transition to phase-locking involves a mixture of LC and multi-level LZ
steps. Next, we calculate the threshold for phase-locking transitions
analytically.

\subsection{Quantum-mechanical ladder climbing}

In the quantum LC regime the nonlinearity parameter \ $P_{2}$ determines the
time interval between successive resonances [see Eq. (\ref{TrMatrix})]. In
the case of a strong nonlinearity, at any given time only two levels are
coupled, and the dynamics can be modeled by successive LZ transitions. In
this case, we can calculate the probability of each transition separately,
and multiply the probabilities. The two level transformation matrix (\ref%
{TrMatrix}) in the energy basis for the $n-1\rightarrow n$ transition yields
the transition probability via the LZ formula \cite{LZ}

\begin{equation}
P_{n-1\rightarrow n}=1-r^{n},  \label{eq:LZ}
\end{equation}%
where $r=e^{-\frac{\pi }{2}P_{1}^{2}}$. We define the probability $P$ for
capture into resonance in this case as the probability of occupying a
sufficiently high energy level $N$ after $N$ successive LZ transitions, i.e.

\begin{equation}
P=\prod_{k=1}^{N}(1-r^{k}),  \label{eq:LC Prob}
\end{equation}%
Then, solving $P(r)=0.5$, one finds the threshold for the LC transition,

\begin{equation}
P_{1}^{cr}=0.79,  \label{eq:Qunt_P1cr}
\end{equation}%
where for two digits accuracy we used $N=5$ in the rapidly converging
product (\ref{eq:LC Prob}). Thus, the capture into resonance occurs in the
first few LZ transitions and one can choose $n_{c}=5$ (see Fig. \ref%
{Flo:P2=8} in the definition Eq. (\ref{eq:Numeric_Prob}) for calculating the
capture probability near the threshold. This prediction is valid for large $%
P_{2}$, as mentioned above. The dashed line in Fig. \ref{Flo:center}
represents Eq. (\ref{eq:Qunt_P1cr}), while the numerical result for 1/2
capture probability is shown by full circles. One can see a very good
agreement between the two results in the LC limit ($P_{2}>5$). However, in
the intermediate regime ($1<P_{2}<5$), oscillations in $P_{1}^{cr}$ are
observed before convergence at the predicted LC line. These oscillations are
due to the mixing of more than two neighboring levels in passage through
resonance (see Fig. \ref{Flo:P2_1}).

\subsection{Classical autoresonance}

As $P_{2}$ decreases, a growing number of levels are coupled simultaneously
and the dynamics becomes increasingly classical. The classical AR phenomenon
is now well understood \cite{Fajans}. If one starts in the zero amplitude
equilibrium, the autoresonant phase-locking is achieved for drives of
amplitude $\varepsilon $ above the critical value $\varepsilon
_{cr}=1.34\alpha ^{3/4}\beta ^{-1/2}m\omega _{0}^{1/2}$ \cite{Fajans}. When
expressed in terms of $P_{1,2}$, this classical threshold is translated into

\begin{equation}
P_{1}^{cr}=0.82/\sqrt{P_{2}}.  \label{eq:Clas_P1cr}
\end{equation}%
When thermal fluctuations are included, the transition probability develops
a width scaling as $T^{1/2}$ with temperature \cite{PRL}. At the same time,
the threshold for 1/2 capture probability remains the same. Thus, $%
P_{1}^{cr} $ in Eq. (\ref{eq:Clas_P1cr}) is the classical counterpart of the
quantum-mechanical observable $P_{1}^{cr}$ in Eq. (\ref{eq:Qunt_P1cr}). This
classical threshold is shown in Fig. \ref{Flo:center} by dashed-dotted line,
illustrating excellent agreement with simulations (full circles) in the
classical regime, $P_{2}\ll 1$. It should be emphasized that the simulation
results in the Figure are solutions of the \emph{quantum-mechanical equations%
} (\ref{eq:slow eq.}) with parameters in the classical regime, while the
probabilities of capture were calculated using the proper transition level $%
n_{c}$ for each value of $P_{2}$, as defined in Eq. (\ref{eq:Numeric_Prob}).
In the next Section, we discuss the width of the autoresonant transition.

\section{The width of the phase-locking transition}

\begin{figure}[tbp]
\includegraphics[width=8cm]{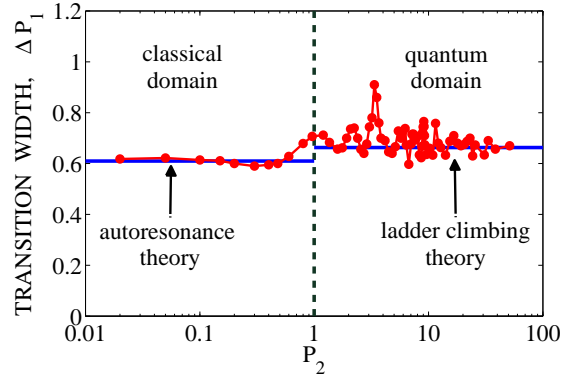}
\caption{(color online) The width of the phase-locking transition in passage
through resonance. Numerical results (full circles) are compared with the
theoretical predictions (solid lines) in the classical AR (on the left) and
the quantum LC (on the right) regimes. The system was in the ground state
initially.}
\label{Flo:width}
\end{figure}

Another observable of the phase-locking transition mentioned above is the
width of the transition, which we define as the inverse slope $(\partial
P/\partial P_{1})^{-1}$ of the phase-locking probability at $P=1/2$. This
width depends on the initial conditions governed by the thermal equilibrium
with the environment. Classically, the thermal width of the autoresonant
transition scales as \cite{PRL}

\begin{equation}
\Delta \varepsilon =1.23\sqrt{\alpha mk_{B}T}.  \label{eq:transition width}
\end{equation}%
However, at very low temperatures, the classical thermal noise becomes
negligible, but quantum fluctuations remain. Recent experiments in Josephson
circuits \cite{Kater} demonstrated quantum saturation of the transition
width at the value obtained from Eq.(\ref{eq:transition width}), but with $%
k_{B}T$ replaced by the energy $\frac{\hbar \omega _{0}}{2}$ of the ground
level. More generally, it was suggested to calculate the width by replacing $%
T$ in the classical formula by an effective temperature, $T_{\mathrm{eff}}=%
\frac{\hbar \omega _{0}}{2k_{B}}\coth (\frac{\hbar \omega _{0}}{2k_{B}T})$,
in agreement with the experimental results. Using $T_{\mathrm{eff}}$, we can
translate Eq. (\ref{eq:transition width}) into the transition width in terms
of $P_{1\text{ }}$

\begin{equation}
\Delta P_{1}^{cl}=1.23\sqrt{k_{B}T_{\mathrm{eff}}/2\hbar \omega _{0}},
\label{A}
\end{equation}%
yielding
\begin{equation}
\Delta P_{1}^{cl}=0.61  \label{B}
\end{equation}%
in the zero temperature limit. The Josephson circuit experiments \cite{Kater}
were performed with $P_{2}=0.00053$, i.e. well inside the classical region
(see Fig. \ref{Flo:center}). Interestingly, these experiments allowed to
characterize the initial quantum "temperature" $T_{\mathrm{eff}}$ of the
system by measuring the final classical autoresonant state of the chirped
excitation. We will justify this approach in the next Section by analyzing
the dynamics of the associated Wigner function in phase space. In contrast
to Eq. (\ref{A}) valid when the final state of the system is classical $%
(P_{2}\ll 1)$, the threshold width of the phase-locking transition in the LC
regime ($P_{2}\gg 1$) can be calculated by evaluating the slope of $P(P_{1})$
from Eq. (\ref{eq:LC Prob}) at $P_{1}=P_{1}^{cr}=0.79$, yielding

\begin{equation}
\Delta P_{1}^{qm}=0.66,
\end{equation}%
where we assume that the system is in the ground state initially. Figure \ref%
{Flo:width} summarizes our theoretical predictions for the width of the
phase-locking transition (for the same parameters as in Fig. \ref{Flo:center}%
) and compares them with those from numerical simulations via the
Schrodinger equation (\ref{eq:slow eq.}). We see a good agreement in both
the AR and LC limits, but notice significant oscillations of the width in
the intermediate range of $P_{2}$. Remarkably, while the thresholds in the
classical and quantum-mechanical limits have very different scalings, the
widths of the transitions are nearly the same.

\section{Chirped dynamics in phase space}

Phase space dynamics comprises a convenient framework for comparison between
classical and quantum evolution of the system. The Wigner function is one of
the most useful phase space representations of the quantum mechanics, since
it reduces to the classical phase space distribution in the limit of $\hbar
\rightarrow 0.$ In this Section, we will study the dynamics of the Wigner
function in our chirped oscillator problem in both the fixed and the
rotating frames and discuss the transition to the classical limit in the
problem.

\subsection{Wigner dynamics in the fixed frame}

The Wigner function $f(x,u,t)$ associated with the $1D$ Hamiltonian of form $%
H(x,p)=\frac{p^{2}}{2m}+V(x,t)$ is governed by the quantum Liouville
equation \cite{Schleich}
\begin{equation}
\frac{\partial f}{\partial t}+u\frac{\partial f}{\partial x}-\frac{1}{m}%
\frac{\partial V}{\partial x}\frac{\partial f}{\partial u}%
=\sum_{l=1}^{\infty }\frac{(-1)^{l}\left( \frac{\hbar }{2m}\right) ^{2l}}{%
m(2l+1)!}\frac{\partial ^{2l+1}V}{\partial x^{2l+1}}\frac{\partial ^{2l+1}f}{%
\partial u^{2l+1}},  \label{eq:Q Lioville}
\end{equation}%
where $u=p/m$ and we neglect possible decay and decoherence processes. We
take a low temperature limit, neglect the nonlinearity initially, and assume
that the initial state of the system is in equilibrium with the environment,
i.e. \cite{Schleich},

\begin{equation}
f_{0}(x,u)=\frac{m\omega _{0}}{2\pi k_{B}T_{\text{eff}}}e^{-\frac{m\omega
_{0}^{2}x^{2}+mu^{2}}{2k_{B}T_{\text{eff}}}},  \label{eq:thermal dist.}
\end{equation}%
where $T_{\mathrm{eff}}=(\hbar \omega _{0}/2k_{B})\coth (\hbar \omega
_{0}/2k_{B}T)$ is the effective temperature. Note that $T_{\mathrm{eff}}$ $%
\rightarrow $ $T$ at high temperatures, while $T_{\mathrm{eff}}\rightarrow
\hbar \omega _{0}/2k_{B}$ at $T$ $\rightarrow 0$.

In the case of interest the potential is a quartic [see Eq. (\ref%
{eq:Hamiltonian})] and, therefore, only one term survives in the right hand
side of (\ref{eq:Q Lioville}), allowing to rewrite this equation in the
following dimensionless form

\begin{equation}
\frac{\partial f}{\partial t}+\overline{u}\frac{\partial f}{\partial
\overline{x}}-\frac{\partial \overline{V}}{\partial \overline{x}}\frac{%
\partial f}{\partial \overline{u}}=\frac{\gamma ^{2}\overline{\beta }%
\overline{x}}{4}\frac{\partial ^{3}f}{\partial \overline{u}^{3}},
\label{eq:dimensionless lioville}
\end{equation}%
where, $\overline{x}=x/L$, $\overline{u}=u/\omega _{0}L$, $L=\sqrt{k_{B}T_{%
\mathrm{eff}}/m\omega _{0}^{2}}$, $\gamma =\hbar \omega _{0}/k_{B}T_{\mathrm{%
eff}}$, $\overline{\beta }=\beta L^{2}$,

\begin{equation}
\overline{V}=\frac{1}{2}\overline{x}^{2}-\frac{1}{4}\overline{\beta }%
\overline{x}^{4}+\overline{\varepsilon }\overline{x}\cos \varphi _{d},
\end{equation}%
and $\overline{\varepsilon }=\varepsilon /mL\omega _{0}^{2}$. In addition,
we measure time $t$ in Eq. (\ref{eq:dimensionless lioville}) in units of $%
\omega _{0}^{-1}$ and introduce the dimensionless chirp rate $\overline{%
\alpha }=\alpha /\omega _{0}^{2}$. With this rescaling, the initial Wigner
distribution (\ref{eq:thermal dist.}) becomes $f_{0}=(2\pi )^{-1}\exp [-(%
\overline{x}^{2}+\overline{u}^{2})/2]$. We solved Eq. (\ref{eq:dimensionless
lioville}) numerically with the same parameters as in the Schrodinger
simulations and show the results in Figs. \ref{Flo:P2=8}-\ref{Flo:P2=0.2}
(subplots e-h) at the same times for comparison. For a better representation
of the Wigner distributions for different nonlinearities, we rescaled the $%
\overline{u},\overline{x}$ axis in the Figures to $\upsilon =\sqrt{\overline{%
\beta }}\bar{u}$, and $\xi =\sqrt{\overline{\beta }}\bar{x}$. The dashed
lines in the Figures are the separatrices, enclosing all bounded classical
trajectories in phase space. We started all these simulations in the ground
state, i.e. $\gamma =2$, at the initial time $\tau _{0}=-8$. Figure \ref%
{Flo:P2=8} compares the dynamics in phase space to that in the energy basis
in the quantum LC regime $(P_{2}=8)$, using the parameters $\overline{\alpha
}=6.25\times 10^{-7}$, $\overline{\beta }=0.0042$, and $\overline{%
\varepsilon }=0.013$. The pattern seen near the origin in Fig. \ref{Flo:P2=8}
is due to the quantum interference with a finite number of states in the
nonresonant region. Figure \ref{Flo:P2_1} shows the intermediate $(P_{2}=1)$
case for parameters $\overline{\alpha }=10^{-4}$, $\overline{\beta }=0.0067$%
, and $\overline{\varepsilon }=0.02$. Finally, Fig. \ref{Flo:P2=0.2}
corresponds to the classical AR case $(P_{2}=0.2)$ and the parameters $%
\overline{\alpha }=10^{-4}$, $\overline{\beta }=0.0013$, and $\overline{%
\varepsilon }=0.038$. As well known \cite{Zurek}, in the near classical case
the Wigner function becomes oscillatory on increasingly fast phase space
scales. However, if coarse-grained (due to a finite numerical accuracy in
our case), the Wigner function becomes almost everywhere positive as one
approaches the classical distribution function, despite the initial
quantum-mechanical ground state used in the simulations. The evolution of
the Wigner function in the last example is nearly classical with the quantum
signature entering only via the effective temperature $\frac{\hbar \omega
_{0}}{2k_{B}}$ of the initial state. In the classical formula (\ref%
{eq:transition width}) for the transition width, $T$ appears due to
integration over the classical Maxwell-Boltzman distribution function (see
\cite{PRL}). Therefore, for the quantum-mechanical initial conditions, we
should integrate over the Wigner function in a thermal state instead over
the classical distribution. But these two distributions have the same
functional shape, except that $T$ is replaced by $T_{\mathrm{eff}}$ in Eq. (%
\ref{eq:thermal dist.}). Therefore, as also confirmed in experiments \cite%
{Kater}, one can use the classical formula for the threshold of the
phase-locking transition at low temperatures, when starting from
quantum-mechanical initial conditions.

\subsection{The dynamics in the rotating frame}

Here we further expand our discussion of the classical AR limit in our
system via the Wigner representation in the rotating frame. The
transformation to the rotating frame is accomplished using unitary
transformation (see \cite{Dykman 2006}) $U=\exp (-i\hat{a}^{\dagger }\hat{a}%
\varphi _{d}),$ where the operator $\hat{a}=(2m\hbar \omega _{0})^{-\frac{1}{%
2}}(m\omega _{0}x+ip)$ and $\varphi _{d}=\int \omega _{d}dt$ is the driving
phase [see Eq. (\ref{eq:Hamiltonian})]. Then, by neglecting rapidly
oscillating terms, the Hamiltonian (\ref{eq:Hamiltonian}) is transformed to

\begin{equation}
\widetilde{H}=U^{\dagger }HU-i\hbar {U}^{\dagger }\dot{U}\approx \frac{\hbar
}{\lambda }\sqrt{\alpha }G,
\end{equation}%
where
\begin{equation}
G=\frac{\tau }{2}(Q^{2}+P^{2})-\frac{1}{4}(Q^{2}+P^{2})^{2}+\mu Q.
\label{eq:slow Hamiltonian}
\end{equation}%
The parameter $\mu =\sqrt{\frac{3\beta }{32\omega _{0}}}\frac{\varepsilon }{%
m\alpha ^{3/4}}=\frac{1}{2}P_{1}\sqrt{P_{2}}$ in the last equation is
familiar from the theory of the classical AR \cite{PRL}, while $\lambda =%
\frac{3\hbar \beta }{8m\sqrt{\alpha }}=\frac{1}{2}P_{2}$ is the
dimensionless Plank constant, entering the commutation relation for the
rescaled variables
\begin{equation}
\lbrack Q,P]=i\lambda .
\end{equation}%
Here $Q=\tilde{L}^{-1}(x\cos \varphi _{d}+\frac{p}{m\omega _{0}}\sin \varphi
_{d})$, $P=\tilde{L}^{-1}(\frac{p}{m\omega _{0}}\cos \varphi _{d}-x\sin
\varphi _{d})$, where $\tilde{L}^{2}=\hbar /(m\omega _{0}\lambda ),$ and the
dimensionless time associated with the dynamics governed by Hamiltonian (\ref%
{eq:slow Hamiltonian}) is $\tau =\sqrt{\alpha }t$.

Next, we write the quantum Liouville equation in the rotating frame (see
Ref. \cite{Dykman 2007} for similar developments for a constant frequency
drive)
\begin{equation}
\frac{\partial f}{\partial \tau }+\frac{\partial G}{\partial P}\frac{%
\partial f}{\partial Q}-\frac{\partial G}{\partial Q}\frac{\partial f}{%
\partial P}=\frac{\lambda ^{2}}{4}\hat{D}f\text{,}  \label{eq:slow wigner}
\end{equation}%
where $\hat{D}=\left( Q\frac{\partial }{\partial P}-P\frac{\partial }{%
\partial Q}\right) \left( \frac{\partial ^{2}}{\partial Q^{2}}+\frac{%
\partial ^{2}}{\partial P^{2}}\right) $. The initial Wigner distribution (%
\ref{eq:thermal dist.}) in the new variables is
\begin{equation}
f_{0}(Q,P)=\frac{1}{2\pi \sigma ^{2}}e^{-\frac{Q^{2}+P^{2}}{2\sigma ^{2}}},
\label{eq: rotatinf frame dist}
\end{equation}%
where $\sigma ^{2}=\lambda \frac{k_{B}T_{\mathrm{eff}}}{\hbar \omega _{0}}=%
\frac{\lambda }{2}\coth (\frac{\hbar \omega _{0}}{2k_{B}T_{\mathrm{eff}}})$.
The left hand side of the Eq. (\ref{eq:slow wigner}) is identical to the
Vlasov equation describing the evolution of a classical distribution of
particles governed by Hamiltonian (\ref{eq:slow Hamiltonian}) without
collisions and self-fields. Hence, as in the fixed frame, after
coarse-graining the fast phase space oscillations of $f$ in the limit $%
\lambda \rightarrow 0$ ($P_{2}\ll 1$), the dynamics in phase space can be
treated classically \cite{Zurek}. Therefore, both the threshold and the
width of the autoresonant transition can be calculated from the classical
theory as illustrated in Figs. \ref{Flo:center} and \ref{Flo:width},
respectively, despite the quantum-mechanical initial conditions in the
problem. In other words, $P_{2}$ is the measure of the classicality of the
phase-locking transition in our chirped oscillator. Furthermore, in the
limit of $P_{2}\ll 1$, only two parameters, $\mu =\frac{1}{2}%
P_{1}P_{2}^{1/2} $ and $T_{\mathrm{eff}}$ $\ $(via the initial conditions)
fully characterize the AR transition. This result is in agreement with Eqs. (%
\ref{eq:Clas_P1cr}) and (\ref{A}) for the AR threshold and its width, where,
remarkably, $\mu $ and $T_{\mathrm{eff}}$ enter separately.

\section{Conclusions}

In conclusion,

(a) We have studied the interrelation between the quantum-mechanical and
classical dynamics of phase-locking transition in a Duffing oscillator
driven by a chirped frequency oscillation. We studied the conditions for a
continuous phase-locking in the driven system, such that the energy of the
oscillator grows to stay in resonance with the varying driving frequency.
The problem was defined by the temperature $T$ and three parameters, i.e.
the driving amplitude $\varepsilon ,$ the driving frequency chirp rate $%
\alpha $, and the parameter $\beta $ characterizing the nonlinearity of the
oscillator. The nonlinearity in the problem was essential, since no
persistent phase-locking in the system could be achieved for $\beta =0$.

(b) We have exploited a more natural representation of both the
quantum-mechanical and classical dynamics in the problem via just two
dimensionless parameters \cite{Gilad}, $P_{1}=\varepsilon /\sqrt{2m\hbar
\omega _{0}\alpha }$ and $P_{2}=(3\hbar \beta )/(4m\sqrt{\alpha }),$ instead
of $\varepsilon ,$ $\alpha $, and $\beta $. \ We have shown that $P_{2}$
describes the classicality of the phase-locking transition in the system,
such that, for $P_{2}\ll 1,$ the system arrives at its classical
autoresonant (AR) state after passage through linear resonance even when
starting in the quantum-mechanical ground state. In contrast, for $P_{2}\gg
P_{1}+1$, the transition involves the energy ladder climbing (LC) process,
i.e. a continuing sequence of separated Landau-Zener transitions between
neighboring energy levels. The parameters $P_{1,2\text{ }}$have a meaning
only in the case of a finite chirp rate, which introduces a new time scale, $%
T_{S}=1/\sqrt{\alpha }$, in the problem.

(c) The probability of transition to the phase-locked state versus $P_{1}$
has a characteristic $S$-shape (a smoothed step function). The value of $%
P_{1}$ yielding 50\% transition probability can be viewed as the threshold
for the phase-locking transition. We have calculated this threshold and its
width in both the quantum-mechanical LC and classical AR limits and compared
the results to those from quantum-mechanical calculations starting in the
ground state of the oscillator (see Figs. \ref{Flo:center} and \ref%
{Flo:width}). We have found that, while in the LC limit the threshold is
independent of $P_{2}$, in the classical AR regime, the threshold is defined
by the combination $\mu =\frac{1}{2}P_{1}P_{2}^{1/2}$ of parameters. The
agreement of the theory and simulations in both limits was excellent, but
characteristic oscillations of the threshold and the width were observed in
the intermediate regime $1<P_{2}<5$.

(d) We have also studied the dynamics of the phase-locking transition in
phase space by using the Wigner function representation, to explain the
quantum saturation of the width of the threshold for AR transitions. The
analysis of the Wigner (quantum Liouville) equation in the chirped rotating
frame clarifies the role of $P_{2}$ as characterizing the degree of
classicality in the phase-locking transition problem.

(d) A possibility of engineering and control of a desired quantum state of
the oscillator via ladder climbing process (see an example in Fig. \ref%
{Flo:P2=8}) seems to be attractive in such applications as quantum
computing. A generalization of this study to include possible decay,
decoherence, and tunneling processes also seems to be important in future
studies.

\begin{acknowledgements}
This work was supported by the Israel Science Foundation under grant No.
451/10.
\end{acknowledgements}

\end{document}